\documentstyle[12pt]{article}
\begin{document}
\baselineskip=15pt

\newcommand{\s}{\mbox{$\sigma$}}
\newcommand{\be}{\begin{equation}}
\newcommand{\ee}{\end{equation}}
\newcommand{\bq}{\begin{eqnarray}}
\newcommand{\eq}{\end{eqnarray}}
\newcommand{\k}{\mbox{$\kappa$}}
\newcommand{\C}{\mbox{$\cal C$}}
\newcommand{\D}{\mbox{$\cal D$}}
\newcommand{\p}{\mbox{$\phi$}}
\newcommand{\r}{\right}
\newcommand{\m}{\left}
\newcommand{\DD}{\Delta}
\newcommand{\del}{\delta}
\newcommand{\T}{\mbox{$\tau$}}
\newcommand{\n}{\mbox{$\bar{\xi}$}}
\newcommand{\vs}{\mbox{$\varsigma$}}
\newcommand{\e}{\mbox{$\epsilon$}}
\newcommand{\vP}{\mbox{$\varphi$}}
\newcommand{\al}{\mbox{$\alpha$}}
\newcommand{\z}{\mbox{$\zeta$}}
\newcommand{\ba}{\mbox{$\beta$}}
\newcommand{\G}{\mbox{$\cal G$}}
\newcommand{\Z}{\mbox{$\cal Z$}}
\newcommand{\Ga}{\mbox{$\Gamma$}}
\newcommand{\Om}{\mbox{$\Omega$}}
\newcommand{\La}{\mbox{$\Lambda$}}
\newcommand{\I}{\mbox{$\cal I$}}
\newcommand{\h}{\mbox{$\theta$}}
\newcommand{\vb}{\mbox{\boldmath${\varphi}$}}

\begin{titlepage}
\rightline{DTP 95/15}
\rightline{hep-th/9504062}
\vskip1in
\begin{center}
{\LARGE Topological Renormalisation}
\end{center}
\begin{center}
{\LARGE of the O(3) Sigma Model}
\end{center}

\vskip1in
\begin{center}
{\large
Richard Costambeys and Paul Mansfield

Department of Mathematical Sciences

University of Durham

South Road

Durham, DH1 3LE, England}

{\it R.G.Costambeys@durham.ac.uk}

{\it P.R.W.Mansfield@durham.ac.uk}
\end{center}
\vskip1in
\begin{abstract}
\noindent We show that the one-instanton sector moduli-space
divergence of the O(3) Sigma Model leads to an unacceptable dependence
of Green's functions on the arbitrary way that the field is split into
a quantum fluctuation about a classical background. Since the
divergence is associated with the degeneration of field configurations
to those of the zero-instanton sector this arbitrariness may be
cancelled by a `topological counter-term' which we construct.
\end{abstract}

\end{titlepage}
%%%%%%%%%%%%%%%%%%%%%%%%%%%%%%%%%%%%%%%%%%%%%%%%%%%%%%%%%%%%%%%%%%%%

\section{\bf Introduction}

Many field theories of physical interest have distinct topological
sectors \cite{Raj}. So, in the semi-classical approximation, Green's
Functions are
reduced to finite dimensional integrals over moduli that parametrise
topologically non-trivial configurations \cite{God+Man}. Typically
these integrals
diverge, so to make sense of them we might introduce a cut-off in moduli
space. How this is done depends on an arbitrary choice of how we
split the field into classical background and quantum fluctuation.
Amplitudes may then appear to be dependent on this choice. Clearly
physical quantities should not depend on this arbitrary choice and so
some procedure is required to ensure that this is the case. The
purpose of this paper is to propose such a procedure.

For example, in the \( O(3) \) \s-model the
configuration space splits up into discrete pieces each characterised by a
topological charge $q$ \cite{AMP}. Within each sector the action is
minimised by
instanton configurations \cite{Bel+Poly}.
The leading semi-classical contribution to
the vacuum expectation value of some operator $\La$
can be written as \cite{FFS}
\be
\langle \La \rangle
={\sum_q}\int \La(a,b,c)\; {{K^q}\over{(q!)^2}}\; e^{-{h_q}(a,b)}
\;{{{d^2}c}\over{\pi(1+|c|^2)^2}}\; {\prod_j}{d^2}{a_j}\;{d^2}{b_j}
\label{eq:ffspf}
\ee
where $a,b,c$ are the moduli and are complex numbers, and $\La(a,b,c)$
is the value of $\La$ evaluated at each instanton configuration. $K$
is a coupling constant.
\be
{{h_q}(a,b)} = -{\sum^{q}_{i<j}}\; {\ln|{a_i}-{a_j}|^2}
-{\sum^{q}_{i<j}}\; {\ln|{b_i}-{b_j}|^2}
+{\sum^{q}_{i,j}}\; {\ln|{a_i}-{b_j}|^2}
\label{eq:ffsact}
\ee
so the partition function diverges as ${a_i}\rightarrow{b_j}$.

We shall show explicitly how simply cutting the integral off at $|a-b|=\e$
makes
this expectation value depend on the way that the field is split
between the quantum piece and the classical background.
This can be encoded into a ``Ward Identity'' analogous to that
associated with BRST invariance in gauge theories \cite{BRS}.

The divergences are associated with classical configurations
degenerating to configurations belonging to
a topological sector with lower instanton number. This
suggests the possibility of modifying the action in the sector
with lower instanton number in such a way as to cancel the
divergence, and hence the dependence on an arbitrary choice of
quantization procedure.

The purpose of this paper is to show that this can be done for the
\( O(3) \) \s-model in the one instanton sector by adding a
``counterterm'' in the
zero instanton sector. This ``renormalisation'' of the topological expansion is
analogous to what occurs in Bosonic String Theory \cite{Paul}.

There are many similarities between the \( O(3) \) \s-model,
Yang-Mills Theory and Bosonic String Theory
which make the \s-model a useful toy model. All three are classically
conformally invariant \cite{D+P}, \cite{C+G}.
The \s-model and Yang-Mills Theory
both display confinement, renormalisability, asymptotic freedom, and instantons
characterised by integer-valued topological indices \cite{Adda},
\cite{BPST} \cite{Jack}.
Other uses of the \( O(3) \) \s-model are well known: i.e. it is
equivalent to the $C{P^1}$ model \cite{Adda} and it is relevant in
describing the
isotropic ferromagnet \cite{Raj}. The \( O(3) \) \s-model is
particularly well-suited to our discussion since the moduli-space
divergence occurs at short distances and so is reliably computed in
the semi-classical approximation.
In Yang-Mills theory and String Theory, gauge-fixing is introduced by
means of the Faddeev-Popov trick \cite{FP}. This may be extended so as to make
the integration over the moduli explicit.
An analogous method may be used in
\s-models, such that the separation of the field into a classical
background and quantum fluctuation may be done by imposing a
constraint on the latter. If this is done with a delta function then
the Faddeev-Popov method generates the appropriate moduli space
measure. The constraint can be included in the action by means
 of ghost-like
fields (which we shall call quasi-ghosts).

So the choice of constraint is important. There are popular choices,
but these are by no means unique. In the development of Polyakov's
formulation of Bosonic String Theory the metric degree of freedom to be
integrated over is set equal to a fiducial metric depending on the
moduli of a Riemann surface \cite{Poly}, \cite{D+P}.
In the semi-classical expansion of Yang-Mills theory \cite{t'H} \cite{Osb}
two constraints
are chosen, one being a background gauge condition, the other
requiring that fluctuations of the gauge-field $A$ about a classical
solution ${\cal A}$ be orthogonal to the derivatives of the latter
with respect to the moduli $\{t^A\}$ \cite{Paul}.
\be
\m[ {\partial_\mu}+{\cal A}_\mu,A_\mu-{\cal A}_\mu \r]=0 \;\;,\;\;
\int {d^4}x\;tr \m({{\partial{\cal A}_\mu}\over{\partial{t^A}}}
\m(A-{\cal A}\r)_\mu\r)=0
\ee
Similarly, a natural, but not unique, choice of condition on
fluctuations of the \( O(3) \) \s-model field, leading to
(~\ref{eq:ffspf}), is to demand that they are orthogonal to the
derivatives of the classical instanton solution with respect to the moduli.

The construction of this paper is as follows. In Section 2 we look at
the Green's Function for a field theory that admits instantons, and
how we can separate the integral over the instanton moduli from that
over the field. Our solution is to use a method analogous to that
devised by Faddeev and Popov introducing a constraint on
a quantum fluctuation about the instanton solution. We also see how
the Green's Function is affected by varying it with respect to this
constraint, finding that this results in an ``anomalous Ward Identity."
In section 3 we review the results of \cite{FFS} which
will be used in this paper. In section 4 we calculate the anomalous
Ward Identity for the case of the one instanton solution in the
\( O(3) \) \s-model. It is found that in order to do the integral over
the instanton moduli we must compactify our solution onto a sphere,
a method for doing this is proposed.

Appendix A contains the detailed calculation for finding the two-point
Green's Function of the fluctuation operator of the \( O(3) \)
\s-model. In Appendix B we describe the one instanton K\"{a}hler
Metric \cite{Nak} for the space of instanton moduli.

\vskip0.5in
%%%%%%%%%%%%%%%%%%%%%%%%%%%%%%%%%%%%%%%%%%%%%%%%%%%%%%%%%%%%%%

\section{\bf General Formulation}

For a field theory in which the configuration space splits into
distinct topological sectors, the partition function can be written as
a sum of partition functions for each sector.
Individual partition functions may be expressed as functional integrals
over a particular homotopy class. So for a theory with fields $w$ and
action $S[w]$
\be
Z=\sum_q {\kappa^q} {Z_q} \;\; , \;\; {Z_q}={\int_{\C_q}}\D w\; {e^{-S[w]}}
\ee
where \(\C_q\) are the distinct homotopy classes, labelled by integers $q$,
of configurations of the
fields \(w\), and \(\k^q\) is some function of the topological
coupling constant.
The Green's Functions are then obtained as sums over the topological
sectors
\bq
{{\G}_q}&=&{\int_{\C_q}}\D w\; {e^{-S[w]}}\;{\La(w)} \\
\G&=&{1\over Z}\; {\sum_q}\;  {\kappa^q} {{\G}_q} \label{eq:GF}
\eq
In each class there is in general a family of
solutions, $v$, to the classical equations of
motion which are parametrized by moduli $\{t^A\}$.
Thus
\be
\m.{{\del S}\over{\del w}}\r|_{w=v} =0 \ \ \ \ \rm{for\;all}\;{t^A}
\ee
Differentiating with respect to $t^A$ shows that
${{{\partial}v}\over{{\partial}{t^A}}}$ is a zero mode of the
fluctuation operator
\be
\m.{{{\del}^2 S}\over{\del w}{\del w}}\r|_{w=v}
\ee
If we set $w=v+\p$ where \p\/
is a quantum fluctuation continuously deformable to zero
then the action may be expanded in powers of \p.
Fluctuations of \p\/ in the directions of the zero modes do not
contribute to the action to leading order and so are not exponentially
damped. They correspond to variations of the moduli. To separate the
integrals over $t$ from the integrals over \p\/, we might try to choose
\p\/ to be orthogonal to the zero modes if this is possible, but this
is arbitrary. More generally, we could introduce some arbitrary
constraint on \p\/ to specify that the fluctuations are transverse to
the space of zero modes. We do this by means of
a set of functions \({F_j}(w,t)=0\), which are introduced into
the Green's Function in a manner analogous to the Faddeev-Popov method,
i.e. by multiplying the Green's Function by
\be
\int {d t}\; {\DD[w,t]}\; {\prod_j}\;\del\m({F_j}(w,t)\r) = 1
\ee
Then we can represent the constraint in the action by using ghosts.
Suppose that the constraints have a solution $w={\hat w}$ and $t={\hat
t}$, then expanding the constraint about $t={\hat t}+{\tilde t}$
\be
{F_j}(w,t)={F_j}({\hat w},{\hat t}) +{{\tilde t}^A}{\partial_A}
{\m.{F_j}(w,t)\r|_{w={\hat w},t={\hat t}}}+ \cdots
\ee
where ${\partial_A}={{\partial}\over{\partial{t^A}}}$. So
\be
\int {d {\tilde t}}\; {\DD[w,t]}\; {\prod_j}\;\del\m({{\tilde t}^A}{\partial_A}
{\m.{F_j}(w,t)\r|_{w={\hat w},t={\hat t}}}\r) = 1
\ee
and we may factor out the operator $\DD$. Integrating out the delta
function leaves the Jacobian, thus
\be
\DD=\det \m({\partial_A}{\m.{F_j}(w,t)\r|_{w={\hat w},t={\hat t}}}\r)
\ee
Representing this determinant using Grassmann numbers we have a
quasi-anti-ghost
\(\xi^j\) for each constraint ${F_j}$, and a quasi-ghost \(\T^A\) for
each parameter $t^A$. Then
\be
\DD = \int d\xi\;d\T\; \exp \m[ -
{\T^A}{\partial_A}\m({\xi^j}{F_j}(w, t)\r) \r]
\ee
Also, if we write
\be
{\prod_j}\;\del\m({F_j}(w,t)\r)=\int {d\lambda}\; e^{-i{\lambda^j}{F_j}}
\ee
then we get
\be
{{\G}_q}=\int {d t}\;{\D W}\; {e^{-S_{tot}}}\;{\La(w)}
\ee
where
\be
S_{tot}=S[w]+{\T^A}{\partial_A}\m({\xi^B}{F_B}(w, t)\r)+i{\lambda^j}{F_j}
\ee
and $W$ denotes $w,\xi,\lambda$. We may write this more usefully by
using a `BRST' transformation. If the transformation is parametrised by
a Grassmann number $\eta$, and acts on $w,\xi,\lambda$ but not the
moduli, then it may be written as ${\del_{\eta}}w=\eta \vs w$, where
the operator
\vs\/ is defined by
\be
\vs{\xi^B}=i{\lambda^B}\;,\; \vs \lambda =0 \;,\; \vs \T=0 \;,\;
\vs w=0 \;,\; {\vs^2}=0 \label{eq:defvarsig}
\ee
So we have
\be
S_{tot}=S[w] + (\vs + {\T^A}{\partial_A})({\xi^j}{F_j}(w, t))
\label{eq:totact}
\ee
Now $\{\vs,{\T^A}{\partial_A}\}=0$ because ${\partial_A}$ acts only on
the moduli whereas $\vs$ does not. Also $\m({\T^A}{\partial_A}\r)^2 =0$
as derivatives with respect to the moduli commute with each other.
Thus $\vs + {\T^A}{\partial_A}$ is also nilpotent. Given that $S[w]$
is independent of the moduli we thus have
\be
(\vs + {\T^A}{\partial_A})S_{tot}=0 \label{eq:1}
\ee
So the action is not `BRST' invariant but
${\vs}S_{tot}=-{\T^A}{\partial_A}S_{tot}$.

We have now achieved our purpose of being able to
write the Green's Function as an integral over the moduli explicitly
\be
{{\G}_q}=\int {d t}\; {g(t)} \;\;,\;\;{g(t)}=\int
{\D W}\; {e^{-S_{tot}}}\;{\La(w)}
\ee
If these integrals diverge then they may be regulated by a cut-off
procedure. However we need to find out if the choice of constraint has
any effect on this. If it does then this is obviously unsatisfactory
and we must do something to compensate. Making an arbitrary variation
of the moduli density with respect to the constraint gives
\bq
\del {g}(t)&=&-\int \D W\;{e^{-S_{tot}}}
(\vs + {\T^A}{\partial_A})({\xi^j}\del{F_j})\;{\La(w)} \nonumber \\
&=&-\int \D W\;(\vs + {\T^A}{\partial_A})\;{e^{-S_{tot}}}\;
({\xi^j}\del{F_j})\;{\La(w)}
\eq
because of (~\ref{eq:1}). But $\int \D W\;\vs\;{e^{-S_{tot}}}\;
({\xi^j}\del{F_j})\;{\La(w)}$ vanishes as $\int \D W\;{e^{-S_{tot}}}\;
({\xi^j}\del{F_j})\;{\La(w)}$ is Grassmann odd. Then
\be
\del {g}(t)=-{\partial_A}\int\D W\;{e^{-S_{tot}}}
{\T^A}({\xi^j}\del{F_j})\;{\La(w)} \label{eq:anom}
\ee

If the integration region of the parameters \(t\) is \(M\) which has
boundary $\partial M$ then we may calculate the variation in the
Partition Function by using Stokes' Theorem in the form
\be
{\int_M} {d^n}t\;{\partial\over{\partial{t^A}}}\m({f^A}(t)\r)
={\int_{\partial M}} d{\Sigma_A}\m({f^A}(t)\r)
\ee
So
\be
\del {{\G}_q}=\del {\int_M} dt\; g(t)
=-{\int_{\partial M}} d{\Sigma_A}\int
\D W\;{\T^A}({\xi^j}\del{F_j}){e^{-S_{tot}}}\;{\La(w)}
\label{eq:totanom}
\ee
If this is non-zero then the Green's Function has aquired a dependence
on the arbitrary choice $F$. This is unacceptable and we must find
some way to cure this problem. Typically the divergencies are
associated with a classical configuration degenerating to one of lower
topology. So configurations on the boundary $\partial M$ may be
approximated by configurations in a different topological sector.
Thus (~\ref{eq:totanom}) may possibly
be cancelled by a
counterterm from another topological sector. We shall see that this
kind of `topological renormalisation' can be done in the case of the
\(O(3)\) \s-model, at least for the simplest kind of divergences.

\vskip0.5in
%%%%%%%%%%%%%%%%%%%%%%%%%%%%%%%%%%%%%%%%%%%%%%%%%%%%%%%%%%%%%%%%%%%%
\section{\bf One loop expansion of the \(O(3)\) \s-model}

In this section we shall review the results of \cite{FFS} (further
details may be found in \cite{Woj}). The action of the
two dimensional static \(O(3)\) \s-model is
\be
S={1\over{2k}} \int {\sum_{a=1}^{3}} {\m({\partial_\mu}{\s^a} \r)^2}
{d^2}x \label{eq:act}
\ee
where there are three scalar fields
\({\s^a}(x,y)\) obey \({\sum_{a=1}^{3}}{\s^a}{\s^a}=\s.\s=1\) and
\(\mu=x,y\). $k$ is the coupling constant.  This constraint can be
included in the action by means of
a Lagrange Multiplier which has on shell value
$l=\s.{{\partial}^2} \s$. Thus the model has
field equations
\be
{{\partial}^2}\s-\m(\s.{{\partial}^2}\s \r)\s=0
\ee
The finite action solutions, ${\s^{(0)}}(x,y)$, tend to a constant
as $x$ and $y$
tend to infinity.
Thus it is possible to
compactify the coordinate space plane $R^2$ on to a sphere which we
shall call $S^{2}_{phys}$. Also the space of fields ${\s^a}$, subject
to $\s.\s=1$, is a spherical surface of unit radius. We shall call
this `internal space' $S^{2}_{int}$.

The finite-action configurations of \s\/ (instantons)
are then mappings of $S^{2}_{phys}$ to $S^{2}_{int}$ and thus can be
classified into homotopy classes. Instanton solutions arise
at the  minima of the action (not zero), i.e. when
$S={{4{\pi}q}\over k}$ where $q$ is the integer valued topological
charge which characterises the homotopy sectors.

It is convenient to stereographically project the spheres $S^{2}_{int}$
and $S^{2}_{phys}$
onto planes. The field on the plane is
\be
w(z,{\bar z})=\frac{{\s^1}+i{\s^2}}{1+{\s^3}}
\ee
where $z=x+iy$ is the complex coordinate on the stereographic
projection of $S^{2}_{phys}$.
In this notation the action (~\ref{eq:act}) becomes
\be
S={4\over k} \int {d^2}\!x\; {{{{\partial_z}{w}}{\partial_{\bar z}{\bar w}}+
{\partial_{\bar z}w}{{\partial_z}{\bar w}}}
\over{(1 + |w|^2)^2}} \label{eq:pact}
\ee
where ${\partial_z}={1\over 2}\m({\partial_x}-i{\partial_y}\r)$ and
${\partial_{\bar z}}={1\over 2}\m({\partial_x}+i{\partial_y}\r)$.
The topological charge on the plane is given by
\be
q={1\over \pi} \int {d^2}\!x\; {{{{\partial_z}{w}}{\partial_{\bar z}{\bar w}}-
{\partial_{\bar z}w}{{\partial_z}{\bar w}}}
\over{(1 + |w|^2)^2}} \label{eq:topch}
\ee
which means that
\be
S={{4{\pi}q}\over k} + {8\over k}\int {d^2}x\; {{\partial_{\bar z}w}
{{\partial_z}{\bar w}}
\over{(1 + |w|^2)^2}} \label{eq:sa}
\ee
and the instantons are given by solutions to ${{\partial_{\bar z}w}}=0$.
Thus we may write the $q$-instanton solution as
\be
w = v = c \m({{{\m(z-{a_1}\r)}{\cdots}{\m(z-{a_q}\r)}}
\over{{\m(z-{b_1}\r)}{\cdots}{\m(z-{b_q}\r)}}}\r)
\ee
Note that the number of poles in the instanton is also equal to the
topological charge. Taking the limit $a \rightarrow b$ in a particular
sector removes one of these poles and so we move to a solution with a
topological charge lower by one.

For the one instanton case
\be
w = v = c \m({{z-a}\over{z-b}}\r)
\ee
the vacuum expectation value given in (~\ref{eq:ffspf}), (~\ref{eq:ffsact})
simplifies
\bq
{{\langle \La \rangle}_{1}}&=&\int \La(a,b,c)\; {K^1}\;{1\over{|a-b|^2}}
\;{{{d^2}c}\over{\pi(1+|c|^2)^2}}\; {d^2}{a}\;{d^2}{b}
\nonumber \\
&=&\int \La(a,b,c)\;{\z_1}(a,b,c)\;
{{d^2}c}\; {d^2}{a}\;{d^2}{b} \label{eq:oivev}
\eq
where ${K^1}={2^6}{e^{\gamma-2}}$, $\gamma=0.5772$ is the Euler
number.
This is calculated in \cite{FFS} by the steepest descent method. A
quantum fluctuation about the instanton solution $v$ is defined and is
chosen to be
orthogonal to the zero modes of the fluctuation operator.
The integrand ${\z_1}(a,b,c)$ is divergent as $a
\rightarrow b$, i.e. as we approach the zero instanton sector from the
one instanton sector.

In the zero instanton case $v=c$ and ${K^0}=1$, so
\be
{{\langle \La \rangle}_{0}}=\int
{d^2}c\;{{\La(c)}\over{\pi(1+|c|^2)^2}}
\ee

\vskip0.5in
%%%%%%%%%%%%%%%%%%%%%%%%%%%%%%%%%%%%%%%%%%%%%%%%%%%%%%%%%%%%%%%%%%%%%%%%%%%

\section{\bf Anomalous Ward Identity}

We shall now calculate the `anomaly' given in (~\ref{eq:totanom}) for the
case of the $O(3)$ \s-model.
If we set $a-b=r$
the one instanton solution becomes
\be
v=c \m(1-{r\over{z-b}}\r)
\ee
then as $|r|$ approaches zero,  $v$ approaches the zero
instanton configuration $v=c$, so we will introduce a moduli-space
cut-off by taking $|r|>\epsilon$.
We take $\e$, $b$ and $c$ as the complex instanton parameters,
defining them collectively as $\{ {t^{\al}}\} = (r, b, c)$ and
$\{ t^{\bar{\al}} \} = ({\bar{r}}, {\bar{b}}, {\bar{c}})$. Note
that $w$ is independent of $\{ t^{\al} \}$.
Now suppose that $w$ differs from $v$ by some quantum correction
$\vP(z,{\bar z})$, then $w=v+\vP$. The quantum correction is constrained
to be orthogonal to  a set of arbitrarily chosen functions $\{\psi\}$
\be
F_{\bar{\al}}\equiv(\psi_{\al},\vP)=0=({\bar\psi}_{\bar{\al}},{\bar{\vP}})
\equiv F_{\al}
\ee
The inner product is defined as
\be
(\psi_{\al},\vP)=\int {d^2}x\; {\sqrt g}\; {\rho^{-2}}\;
{\bar{\psi}}_{\bar{\al}}\; \vP \;\;,\;\; \rho=1+{|v|}^2 \label{eq:innprod}
\ee
where ${g_{\mu\nu}}$ is the metric on  $S^{2}_{phys}$, and $g=\det
{g_{\mu\nu}}$.
The action (~\ref{eq:sa}) may be expanded in orders of $\vP$. To
leading order it is
\be
{S_0}[\vP]={{4{\pi}q}\over k} + {8\over k}(\vP,\DD\vP)
\ee
where the fluctuation operator is
\be
\DD=-{{\rho^2}\over{\sqrt g}}{\partial_z}{\rho^{-2}}{\partial_{\bar
z}}
\ee
where ${\partial_z}={{\partial}\over{\partial{\bar z}}}$.
This is different to the fluctuation operator used in \cite{FFS}, but this
is just because the variables in the second term of the action are
defined differently in each case. It is easy to show that the
fluctuation operators are equivalent.

The functionals \(F_A\) are the constraints which enter into the action in the
manner outlined in the Section 2. Therefore the total
action (~\ref{eq:totact}) is
\be
S_{tot}={S_0}[\vP] + (\vs + {\T^A}{\partial_A})({\xi^B}{F_B})
\ee
and the `anomaly' (~\ref{eq:totanom})
\be
\del {{\G}_1}=-{\int_{\partial M}} d{\Sigma_A} \int\D \Phi \;{e^{-S_{tot}}}
{\T^A}({\xi^B}\del{F_B})\;{\La(v)}
\ee
where $\Phi$ denotes all the fields.
Looking to integrate out the ghosts we expand out the action using
the properties of the operator $\vs$ given in (~\ref{eq:defvarsig})
\bq
S_{tot}&={S_0}[\vP] + {{\lambda}^A}{F_A} &
- {\T^{\al}}{{\xi}^{\bar{\ba}}}m_{\al{\bar{\ba}}}
-{\T^{\bar{\al}}}{{\xi}^{\ba}}{\bar{m}}_{{\bar{\al}}\ba} \nonumber \\
& &+ {\T^{\al}}{{\xi}^{\ba}}
(\vP,{\partial_{\al}}{\psi}_{\ba})
+ {\T^{\al}}{{\xi}^{\bar{\ba}}}
({\bar{\vP}},{\partial_{\al}}{\bar{\psi}}_{\bar{\ba}}) \nonumber \\
& &+ {\T^{\bar{\al}}}{{\xi}^{\ba}}
(\vP,{\partial_{\bar{\al}}}{\psi}_{\ba})
+ {\T^{\bar{\al}}}{{\xi}^{\bar{\ba}}}
({\bar{\vP}},{\partial_{\bar{\al}}}{\bar{\psi}}_{\bar{\ba}}) \label{eq:22}
\eq
The ghost field propagators have been written as
\be
m_{\al{\bar{\ba}}}=\m({\partial_{\bar{\al}}}{\bar v},
{\bar{\psi}}_{\bar{\ba}}\r)\;\;,\;\;
{\bar{m}}_{{\bar{\al}}\ba}=\m({\partial_{\al}}v,{\psi}_{\ba}\r)
\label{eq:defm}
\ee
The last four terms of (~\ref{eq:22}) are
the interaction terms between the ghost fields and $\vP$.
Now let ${\tilde{S}}= {S_0} +{{\lambda}^A}{F_A}- {\T^{\al}}
{{\xi}^{\bar{\ba}}}m_{\al{\bar{\ba}}}
-{\T^{\bar{\al}}}{{\xi}^{\ba}}{\bar{m}}_{{\bar{\al}}\ba}$
and expand the rest of
the action in a power series in $\vP$. So if ${S_{tot}}={\tilde{S}}+{\bar{S}}$
then
\be
e^{-{\bar{S}}} = 1 -
\m[{\T^M}{{\xi}^\nu}
({\vP},{\partial_M}{\psi}_{\nu})+
{\T^M}{{\xi}^{\bar{\nu}}}
({\bar{\vP}},{\partial_M}{\bar{\psi}}_{\bar{\nu}})
\r]+\cdots
\ee
$M$ stands for both $\mu$ and ${\bar{\mu}}$.
The first term in the expansion disappears under contraction with
the rest of ${{\G}_1}$ as it is linear in $\vP$.
The terms to leading order in the coupling are
\bq
\del {{\G}_1}&=&
-{\int_{\partial M}} d{\Sigma_{\al}}\int \D\Phi\; e^{-{\tilde{S}}}
\m({\T^{\al}}{{\xi}^{\ba}}
({\vP},\del {\psi}_{\ba}){\T^{\bar\mu}}{{\xi}^{\bar{\nu}}}
({\bar{\vP}},{\partial_{\bar\mu}}{\bar{\psi}}_{\bar{\nu}})\;{\La(v)}\r)
\nonumber \\
& &\mbox{}{\hskip1in}
-{\int_{\partial M}} d{\Sigma_{\bar{\al}}}\int \D\Phi\; e^{-{\tilde{S}}}
\m({\T^{\bar{\al}}}{{\xi}^{\ba}}({\vP},
\del {\psi}_{\ba})
{\T^{{\mu}}}{{\xi}^{\bar{\nu}}}
({\bar{\vP}},{\partial_\mu}{\bar{\psi}}_{\bar{\nu}})\;{\La(v)}
\r) \nonumber \\
&=&-{\int_{\partial M}} d{\Sigma_{\al}}\m({\z_1}(t){\bar m}^{-1}_{{\bar\mu}\ba}
({\vb},\del {\psi}_{\ba})
{m^{-1}_{\al{\bar\nu}}}
({\bar{\vb}},{\partial_{\bar\mu}}{\bar{\psi}}_{\bar{\nu}})\;{\La(v)}
\r)\nonumber \\
& &\mbox{}{\hskip1in}
-{\int_{\partial M}} d{\Sigma_{\bar{\al}}}\m({\z_1}(t)
{\bar{m}}^{-1}_{{\bar{\al}}\ba}
({\vb},\del {\psi}_{\ba})
{m^{-1}_{\mu{\bar{\nu}}}}
({\bar{\vb}},{\partial_\mu}{\bar{\psi}}_{\bar{\nu}})\;{\La(v)}\r)\nonumber \\
&=&-{\int_{\partial M}} d{\Sigma_{\al}}{{\Psi}^{\al}}\;{\La(v)}
-{\int_{\partial M}} d{\Sigma_{\bar{\al}}}{{\Psi}^{\bar{\al}}}\;{\La(v)}
\eq
Where ${\z_1}(t)$ is the one-loop partition function given above in
(~\ref{eq:oivev}),
and the boldface type indicates that the field $\vb$ is contracted with
its conjugate field in the same term.
We will now show that ${\Psi^{\bar{\al}}}$ and ${\Psi^{\al}}$ can
themselves be
expressed as variations. This enables us to both simplify their
evaluations and find counterterms which will cancel $\del {\int_M}
dt\; g(t)$.
To do
this we see that the fields $\vb$ contract to introduce a two-point
Greens function. The Greens function
of the two-point fluctuation operator is found in Appendix A to be
\bq
\I(x,y)&=&<{\vP}(x)\;{\bar{\vP}}(y)> =\vb(x) {\bar{\vb}}(y) \\
&=&-{1\over{\pi^2}}\int d^2 z \;
(1-P^{\dag}.){1\over{x-z}}{\rho^2}(z)
{1\over{{\bar{z}}-{\bar{y}}}}(1-.P)
\eq
here we have used a dot notation equivalent to the inner product,
i.e.
\be
P^{\dag}(y,x).{1\over{x-z}}=\int d^2 {x^{\prime}}\; P^{\dag}(y,{x^{\prime}})
\;{1\over{{x^{\prime}}-z}}\; {\sqrt g}\;{\rho^{-2}}({x^{\prime}})
\ee
where
\be
P(x,y)={\psi_a}(x){{\bar{m}}^{-1}_{{\bar{b}}a}}{{\bar{\Z}}_{\bar{b}}}(y)
\ee
\be
P^{\dag}(y,x)={{\Z}_{b}}(y){m^{-1}_{b{\bar a}}}{{\bar{\psi}}_{\bar
a}}(x)
\ee
and $\Z(x)$ is a zero mode of $\DD$. Above we showed that
the zero modes of the fluctuation operator are ${\partial_A}v$. If
we denote these by
$\Z(x)$ then from (~\ref{eq:defm}) we see that
\be
{m_{{\al}{\bar{\ba}}}}=\m({{\bar{\Z}}_{\bar{\al}}},
{{\bar\psi}_{\bar{\ba}}}\r)
\;\;,\;\;
{\bar{m}}_{{\bar{\al}}\ba}=\m({\Z_{\al}},{\psi}_{\ba}\r)
\ee
We are interested in $\Psi$ evaluated on the cut-off boundary
$\partial M$ given by $|a-b|=|r|=\epsilon$. We show in Appendix A that here
the Green's Function $\I(x,y)$ can be expressed in terms of the zero
instanton sector Green's Function as
\be
\I(x,y)=(1-P^{\dag}.)\;{\I_0}(x,y)\;(1-.P) \times \m(1+O(\epsilon)\r)
\ee
If we denote ${\I_0}(x,y)$ by the zero instanton sector contraction
${{\bar{\vb}}_0}(x) \;\;{{\vb}_0}(y)$ we can approximate $\Psi$ in
terms of the zero instanton sector as follows.
Applying the appropriate parts of
${\Psi^{\al}}$ to $\I(x,y)$
and again using the dot notation
\be
({\varphi},\del{\psi}_{\ba})=
\int {d^2}x\; {\sqrt g}\;
{\rho^{-2}}\;{\bar{\varphi}}\;\del{\psi}_{\ba} =
{\bar{\varphi}}\cdot\del{\psi}_{\ba}
\ee
gives for $f$ an arbitrary function independent of $\psi$
\bq
{\bar{m}}^{-1}_{{\bar\mu}\ba}\m(\m(f(1-\cdot P)\r)\cdot\del{\psi}_{\ba}\r)
&=&{\bar m}^{-1}_{{\bar\mu}\ba}\;\m(f\cdot\del{\psi}_{\ba}\r)
- {\bar m}^{-1}_{{\bar\mu}\ba}\;\m(f\cdot{\psi_a}\r){{\bar m}^{-1}_{{\bar b}a}}
\m({{\bar{\Z}}_{\bar b}}\cdot\del{\psi}_{\ba}\r) \nonumber \\
&=&{\bar m}^{-1}_{{\bar\mu}\ba}\;\m(f \cdot\del{\psi}_{\ba}\r)
- {\bar m}^{-1}_{{\bar\mu}\ba}\;\m(f \cdot{\psi_a}\r){{\bar m}^{-1}_{{\bar
b}a}}
\del{{\bar m}_{{\bar b}\ba}} \nonumber \\
&=&{\bar m}^{-1}_{{\bar\mu}\ba}\;\m(f \cdot\del{\psi}_{\ba}\r)
- \m(f \cdot{\psi_a}\r)\del{{\bar m}^{-1}_{{\bar\mu}a}} \nonumber \\
&=&\del(f \cdot{\psi_{\ba}}{{\bar m}^{-1}_{{\bar\mu}\ba}})
\label{eq:conv1}
\eq
and
\bq
{m^{-1}_{\al{\bar\nu}}} \m(\m((1-P^{\dag}\cdot)f\r)\cdot
{\partial\over{\partial{t^{\bar\mu}}}}{\bar\psi}_{\bar\nu}\r)
&=&{m^{-1}_{\al{\bar\nu}}} \m[{\partial\over{\partial{t^{\bar\mu}}}}
\m(\m((1-P^{\dag}\cdot)f\r)\cdot{\bar\psi}_{\bar\nu}\r) \r.\nonumber \\
& &\mbox{}{\hskip1in}- \m.\m(\m({\partial\over{\partial{t^{\bar\mu}}}}
\m((1-P^{\dag}\cdot)f\r)\r)\cdot
{\bar\psi}_{\bar\nu}\r)\r] \nonumber \\
&=&{m^{-1}_{\al{\bar\nu}}}
\m({{\partial P^{\dag}}\over{\partial{t^{\bar\mu}}}}\cdot f\r)
\cdot{\bar\psi}_{\bar\nu} \nonumber \\
&=&{m^{-1}_{\al{\bar\nu}}}
\m({{\partial}\over{\partial{t^{\bar\mu}}}}\m({\Z_b}{m^{-1}_{b{\bar a}}}
{\bar\psi}_{\bar a}\r)\cdot f\r)\cdot{\bar\psi}_{\bar\nu} \nonumber \\
&=&{m^{-1}_{\al{\bar\nu}}}{m_{b{\bar\nu}}}
\m({{\partial}\over{\partial{t^{\bar\mu}}}}\m({m^{-1}_{b{\bar a}}}
{\bar\psi}_{\bar a}\r)\cdot f\r) \nonumber \\
&=&{\partial\over{\partial{t^{\bar\mu}}}}{m^{-1}_{\al{\bar a}}}
{\bar\psi}_{\bar a}\cdot f
\label{eq:conv2}
\eq
Similarly
\bq
{{\bar m}^{-1}_{{\bar{\al}}\ba}}
\m(\m(f(1-.P)\r).\del{\psi}_{\ba}\r)&=&
\del(f.{\psi_{\ba}}{{\bar m}^{-1}_{{\bar{\al}}\ba}}) \label{eq:conv3}\\
{m^{-1}_{\mu{\bar\nu}}}
\m(\m((1-P^{\dag}.)f\r).
{\partial\over{\partial{t^{\mu}}}}{\bar\psi}_{\bar\nu}\r)
&=&{\partial\over{\partial{t^{\mu}}}}{m^{-1}_{{\bar a}\mu}}{\bar\psi}_{\bar
a}.f\label{eq:conv4}
\eq
so that
\bq
{\Psi^{\al}}&=&{\z_1}(t)\del\m[{{\bar m}^{-1}_{{\bar\mu}\ba}}
({\vb_0},{\psi}_{\ba})\r]
{\partial\over{\partial{t^{\bar\mu}}}}({m^{-1}_{{\bar a}\al}}
{\bar{\vb}_0},{\bar\psi}_{\bar a}) \nonumber \\
&=&\del\m[{\z_1}(t){{\bar
m}^{-1}_{{\bar\mu}\ba}}({\vb_0},{\psi}_{\ba})
{\partial\over{\partial{t^{\bar\mu}}}}({m^{-1}_{{\bar a}\al}}
{\bar{\vb}_0},{\bar\psi}_{\bar a})\r] \label{eq:(32)}
\eq
and
\bq
{\Psi^{\bar{\al}}}&=&{\z_1}(t)\del\m[{{\bar m}^{-1}_{{\bar{\al}}\ba}}
({\vb_0},{\psi}_{\ba})\r]
{\partial\over{\partial{t^{\mu}}}}({m^{-1}_{{\bar
a}\mu}}{\bar{\vb}_0},
{\bar\psi}_{\bar a}) \nonumber \\
&=&\del\m[{\z_1}(t){{\bar
m}^{-1}_{{\bar{\al}}\ba}}({\vb_0},{\psi}_{\ba})
{\partial\over{\partial{t^{\mu}}}}({m^{-1}_{{\bar a}\mu}}{\bar{\vb}_0},
{\bar\psi}_{\bar a})\r] \label{eq:(33)}
\eq
Thus we have succeeded in writing ${\Psi^{\al}}$ and
${\Psi^{\bar{\al}}}$ on $\partial M$ as variations, given in terms of
the zero instanton sector Green's Function. We will now write this as
a zero instanton sector expectation value. In this sector
\be
{\int_0} \D \Phi\;{e^{-S_{tot}}}\;{\bar{\vP}_0}(x) {\vP_0}(y) =
\int {K^0}\;{\bar{\vb_0}}(x) {\vb_0}(y)\;{{d^2 c}\over{\pi
(1+|c|^2)^2}}
\ee
to leading order in the coupling. The one instanton moduli are $b$,
$c$ and $r=a-b$ whose magnitude is held fixed on $\partial M$ at $|r|=\e$,
so we can set
\be
(d{\Sigma^A})=(d{\Sigma^{r}},d{\Sigma^{\bar{r}}}) =
(d{\bar{r}}\;{d^2}b\;{d^2}c\;,\;-d{{r}}\;{d^2}b\;{d^2}c)
\ee
and as ${\z_1}(t)={{K^1}\over{\pi {|r|^2} (1+|c|^2)^2}}$
hence
\bq
\del{{\cal G}_1}&=& \del {\int_0} \D \Phi\;{e^{-S_{tot}}}
\m({\oint_{|r|=\epsilon}} d{\bar{r}}\; {d^2}b\; {{K^1}\over{|r|^2}}
\m({{\bar m}^{-1}_{{\bar\mu}\ba}}({\vP_0},{\psi}_{\ba})
{\partial\over{\partial{t^{\bar\mu}}}}({m^{-1}_{{\bar a}r}}
{\bar{\vP}_0},{\bar\psi}_{\bar a})\r)\r. \nonumber \\
& &\m.\mbox{}-{\oint_{|r|=\epsilon}}
d{{r}}\; {d^2}b\; {{K^1}\over{|r|^2}}
\m({{\bar m}^{-1}_{{\bar{r}}\ba}}({\vP_0},{\psi}_{\ba})
{\partial\over{\partial{t^{\mu}}}}({m^{-1}_{{\bar a}\mu}}{\bar{\vP}_0},
{\bar\psi}_{\bar a})\r)\r){\La(c)}  \\
&\equiv&\del {\int_0} \D \Phi\;{e^{-S_{tot}}}\;{\cal J}\;{\La(c)}
\eq
where
\bq
{\cal J}&=&\int {d^2}b\;\m[
{\oint_{|r|=\epsilon}} d{\bar{r}}\; {{K^1}\over{|r|^2}}
\m({{\bar m}^{-1}_{{\bar\mu}\ba}}({\vP_0},{\psi}_{\ba})
{\partial\over{\partial{t^{\bar\mu}}}}({m^{-1}_{{\bar a}r}}
{\bar{\vP}_0},{\bar\psi}_{\bar a})\r)\r. \nonumber \\
& &\m.\mbox{}-{\oint_{|r|=\epsilon}}
d{{r}}\; {{K^1}\over{|r|^2}}
\m({{\bar m}^{-1}_{{\bar{r}}\ba}}({\vP_0},{\psi}_{\ba})
{\partial\over{\partial{t^{\mu}}}}({m^{-1}_{{\bar a}\mu}}{\bar{\vP}_0},
{\bar\psi}_{\bar a})\r)\r]
\label{eq:J1}
\eq
in terms of the zero instanton sector fields. Since we have succeeded
in writing the variation of the one instanton sector contribution as
the variation of a zero instanton contribution, it is clear that we
can cancel the divergence from this variation by a counterterm in the
zero instanton sector. We will thus modify (~\ref{eq:GF}) by defining
\be
{\tilde{\G}}={1\over {\tilde Z}}\; {\sum_q}\;  {\kappa^q} {\tilde{\G}_q}
\label{eq:greensfn}
\ee
where
\bq
{\tilde{\G}_0}&=&\int \D w\; {e^{-S[w]}}\;{\La(w)}\;(1-{\kappa}{\cal
J}) \\
{\tilde{Z}_0}&=&\int \D w\; {e^{-S[w]}}\;(1-{\kappa}{\cal J}) \\
{\tilde{\G}_1}&=&{\G_1}\;\;,\;\;{\tilde{Z}_1}={Z_1}
\eq
so that to the order that we are working
\be
{\del_{\psi}}{\tilde{\G}}=0
\ee
Clearly to higher orders there are further modifications. We will now
evaluate $\cal J$ for the natural choice of the constraint $F$ which
is when the ${\psi}$ are the zero modes
${\partial{v}\over{\partial{t}}}$ of the fluctuation operator.
In this case the propagators of the
ghost fields (~\ref{eq:defm}) become the K\"{a}hler metric
which is
the metric tensor on the manifold of the instantons, parametrized by
the \(\{t^{\al}\}\).
\(m_{\al{\bar{\ba}}}\) is K\"{a}hler as it can be written in the
form \cite{Nak},\cite{Woj}
\be
m_{\al{\bar{\ba}}}={\partial\over{\partial{t^{\al}}}}
{{\partial}\over{\partial{t^{\bar{\ba}}}}}\; {\cal K}\;\;,\;\;
{\cal K}=\int\;{d^2}x\;{\sqrt{g(x)}}\;\ln(1+|v|^2)
\ee
where ${\cal K}$ is the K\"{a}hler Potential.
\(m_{\al{\bar{\ba}}}\) can be calculated for the one
instanton case (see Appendix B). Henceforth \(m_{\al{\bar{\ba}}}\) shall
mean the K\"{a}hler metric.
To calculate $\cal J$ we have
for $v=c \m(1-{r\over{z-b}}\r)$
\be
{\psi}_{\ba}={{\partial v}\over{\partial{t^{\ba}}}}
={\m(\begin{array}{c} {{\partial v}\over{\partial{r}}}\\ \\
 {{\partial v}\over{\partial{b}}}\\ \\{{\partial v}\over{\partial{c}}}
\end{array} \r)}_\beta
={\m(\begin{array}{c} -{c\over{z-b}}\\ \\  -{{cr}\over{(z-b)^2}}\\ \\
1-{r\over{z-b}} \end{array} \r)}_\beta
\ee
and ${m^{-1}_{{\bar{\al}}\ba}}$ is the inverse of the Kahler metric
\be
\left({m^{-1}_{{\bar{\al}}\ba}}\right)={1\over{m}} \left(\matrix{A&B&C\cr
                                           {B^\dagger}&D&E\cr
                                        {C^\dagger}&{E^\dagger}&F\cr}\right)
\equiv \left({{{m^{\prime}_{{\bar{\al}}\ba}}}\over{m}}\right)
\label{eq:invm}
\ee
where $A$, $B$, $C$, $D$, $E$ and $F$, are all
functions of the $\{ t^{\al}\}$ (see Appendix B),
and $m=\det({m_{{\bar{\al}}\ba}})$.
Also
\be
\left({{\bar m}_{{\bar{\al}}\ba}}^{-1}\right)=
{{\m({m^{\prime}_{{\bar{\al}}\ba}}\r)^{T}}\over{\bar m}}
\ee
both terms in $\cal J$ may
be split into a couple of inner products which may be treated
separately. We need to calculate
\be
{{\bar m}^{-1}_{{\bar\mu}\ba}}\;({{\vP}_0},{\psi}_{\ba})
=\int d^2 z {\sqrt g}\; {\rho^{-2}}\;{{\bar m}^{-1}_{{\bar\mu}\ba}}\;
{\psi}_{\ba}\;{\bar{\vP}_0}(z,{\bar z})
\ee
where
\be
{\rho^{-2}}={1\over{(1+|v|^2)^2}}
={{|z-b|^4}\over{(|z-b|^2 + |c|^2 |z-b-r|^2)^2}}
\ee
and the square root of the determinant of the metric on  $S^{2}_{phys}$ is
\be
{\sqrt g}={1\over{(1+|z|^2)^2}}
\ee
So
\bq
{{\bar m}^{-1}_{{\bar\mu}\ba}}\;({{\vP}_0},{\psi}_{\ba})
&=&{{\m({m^{\prime}_{{\bar{\al}}\ba}}\r)^{T}}\over{\bar m}}
\int d^2 z\; {{|z-b|^4}\over{{(1+|z|^2)^2}(|z-b|^2 + |c|^2
|z-b-r|^2)^2}}\;\nonumber \\
& &\mbox{\hskip1.5in}\times
\pmatrix{-{c\over{z-b}}\cr \cr  -{{cr}\over{(z-b)^2}}\cr  \cr
1-{r\over{z-b}}\cr}_\beta{\bar{\vP}_0}(z,{\bar z}) \label{eq:77}
\eq
However this needs to be evaluated on the plane so that we are able to
use the results of \cite{FFS} which are all given on the plane. To do
this we simply replace ${g}$ with the flat space metric.
These three integrals don't need to be evaluated until later. However
we notice at this stage that
one of the integrals is zero for small $r$
\be
\int d^2 z\;
{{|z-b|^4  {\bar{\vP}_0}(z,{\bar z})}
\over{(|z-b|^2 + |c|^2 |z-b-r|^2)^2}}
\simeq\int d^2 z\; {{{\bar{\vP}_0}(z,{\bar z})}\over{{(1+|c|^2)^2}}} =0
\ee
as this is $\langle1\vert{\bar{\vP}}_0\rangle = 0$, i.e.
${\bar{\vP}}_0$ is orthogonal
to the constant zero mode of $\DD$.
Thus we shall represent (~\ref{eq:77}) by
\bq
{{\bar m}^{-1}_{{\bar\mu}\ba}}\;({{\vP}_0},{\psi}_{\ba})
&=&-{{\m({m^{\prime}_{{\bar{\al}}\ba}}\r)^{T}}\over{\bar m}}
\int d^2 z\;\pmatrix{c\Om \cr cr\Ga \cr r\Om \cr}_\beta
{\bar{\vP}_0}(z,{\bar z}) \nonumber \\
&=&-{1\over{{\bar m} }} \int d^2 z\;
\pmatrix{{c{A}\Om}+r{c{B^\dagger}\Ga + r{C^\dagger}\Om} \cr
{c{B}\Om}+r{c{{D}}\Ga + r{E^\dagger}\Om} \cr
{c{C}\Om}+r{c{E}\Ga+r{{F}}\Om}\cr}_\beta{\bar{\vP}_0}(z,{\bar z})
\label{eq:part1}
\eq
where $\Om$ and $\Ga$ are functions of $z$, $b$ and $c$.
Similarly for
the other inner product we need to calculate from (~\ref{eq:J1})
\bq
({m^{-1}_{{\bar a}\al}}{\bar{\vP}_0},{\bar\psi}_{\bar a})
&=&{{{m^{\prime}_{{\bar{\al}}\ba}}}\over{m}}
\int d^2 w\; {{|w-b|^4}\over{{(1+|w|^2)^2}(|w-b|^2 + |c|^2
|w-b-r|^2)^2}}\; \nonumber \\
&&\mbox{}{\hskip1.5in}\times
\pmatrix{-{{\bar c}\over{{\bar w}-{\bar b}}}\cr \cr
-{{{\bar c}{\bar{r}}}\over{({\bar w}-{\bar b})^2}}\cr  \cr
1-{{\bar{r}}\over{{\bar w}-{\bar b}}}\cr}_\beta{\vP_0}(w,{\bar w}) \nonumber
\eq
we shall represent this as
\bq
({m^{-1}_{{\bar a}\al}}{\bar{\vP}_0},{\bar\psi}_{\bar a})
&=&{{{m^{\prime}_{{\bar{\al}}\ba}}}\over{m}} \int d^2 w\;
\pmatrix{{\bar c}{{\Om}^\prime} \cr {\bar c}{\bar{r}}{{\Ga}^\prime}
\cr {\bar{r}}{{\Om}^\prime} \cr}_\beta {\vP_0}(w,{\bar w}) \nonumber \\
&=&-{1\over{m}} \int d^2 w\;
\pmatrix{ {{\bar c}A{{\Om}^\prime}}+{\bar{r}}{{\bar
c}B{{\Ga}^\prime}+{\bar{r}}C{{\Om}^\prime}} \cr
{{\bar c}{B^\dagger}{{\Om}^\prime}}+{\bar{r}}{{\bar c}{D}{{\Ga}^\prime}
+{\bar{r}}E{{\Om}^\prime}}\cr
{{\bar c}{C^\dagger}{{\Om}^\prime}}+{\bar{r}}{{\bar c}{E^\dagger}
{{\Ga}^\prime}
+{\bar{r}}{F_2}{{\Om}^\prime}}\cr}_\beta
{\vP_0}(w,{\bar w})
\label{eq:part2}
\eq
where $\Om^\prime$ and $\Ga^\prime$ are also functions of $z$, $b$ and $c$.
So using (~\ref{eq:part1}) and (~\ref{eq:part2}) then $\cal J$ is
\bq
{\cal J}&=&\int {d^2}b\;\m[
{\oint_{|r|=\epsilon}} d{\bar{r}}\; {{K^1}\over{|r|^2}}
\int \!d^2 z\!\!  \int \!d^2 w\; {{\cal Y}^{r}}(z,w,r,b,c)
{\bar{\vP}_0}(z,{\bar z}){\vP_0}(w,{\bar w})\r. \nonumber \\
& &\m.\mbox{}-{\oint_{|r|=\epsilon}}
d{{r}}\; {{K^1}\over{|r|^2}}
\int \!d^2 z\!\!  \int \!d^2 w\; {{\cal Y}^{\bar{r}}}(z,w,r,b,c)
{\bar{\vP}_0}(z,{\bar z}){\vP_0}(w,{\bar w})\r]
\eq
where
\bq
{{\cal Y}^{r}}&=&
{1\over{{\bar m} }}
\pmatrix{{c{A}\Om}+r{c{B^\dagger}\Ga + r{C^\dagger}\Om} \cr
{c{B}\Om}+r{c{{D}}\Ga + r{E^\dagger}\Om} \cr
{c{C}\Om}+r{c{E}\Ga+r{{F}}\Om}\cr}
\m({{\partial\over{\partial{\bar{r}}}}},
{{\partial\over{\partial{\bar{b}}}}},
{{\partial\over{\partial{\bar{c}}}}}\r)
{1\over{m}}
\m({{\bar c}A{{\Om}^\prime}}+{\bar{r}}{{\bar
c}B{{\Ga}^\prime}+{\bar{r}}C{{\Om}^\prime}}\r) \\
{{\cal Y}^{\bar{r}}}&=&
{1\over{{\bar m} }} \m({c{A}\Om}+r{c{B^\dagger}\Ga + r{C^\dagger}\Om}\r)
\m({{\partial\over{\partial{r}}}},
{{\partial\over{\partial{b}}}},
{{\partial\over{\partial{c}}}}\r){1\over{m}}
\pmatrix{ {{\bar c}A{{\Om}^\prime}}+{\bar{r}}{{\bar
c}B{{\Ga}^\prime}+{\bar{r}}C{{\Om}^\prime}} \cr
{{\bar c}{B^\dagger}{{\Om}^\prime}}+{\bar{r}}{{\bar c}{D}{{\Ga}^\prime}
+{\bar{r}}E{{\Om}^\prime}}\cr
{{\bar c}{C^\dagger}{{\Om}^\prime}}+{\bar{r}}{{\bar c}{E^\dagger}
{{\Ga}^\prime}
+{\bar{r}}{F_2}{{\Om}^\prime}}\cr}
\eq
To do the boundary integrals over $r$ and $\bar{r}$ we set
$r=\epsilon{e^{i\theta}}$ and ${\bar{r}}=\epsilon{e^{-i\theta}}$ so
$dr=i \epsilon
{e^{i\theta}}d\theta$ and $d{\bar{r}}=-i{{\epsilon}}{e^{-i\theta}}
d\theta$. Only the $\theta$-independent terms in the integrands
contribute to $\oint d\theta {1\over r} {{\cal Y}^{r}}$ and
$\oint d\theta {1\over{\bar{r}}} {{\cal Y}^{\bar{r}}}$. There is
only one divergent piece from each
term, and these pieces turn out to be the same. For
$\oint d\theta {1\over r} {{\cal Y}^{r}}$ we find
\be
{1\over{r{\bar m}}}\m(c{A_1}\Om\r){{\partial }\over{\partial{\bar{r}}}}
\m({1\over{m}}\m({\bar c}{A_1}{\Om^{\prime}}\r)\r)
={{{S^4} {\Om}{{\Om}^\prime}}
\over{{|c|^2}{\epsilon^2}{(ln({\epsilon^2}))^3}}} \label{eq:divterm1}
\ee
where $A_1$ and $m$ are given in Appendix B.
For $\oint d\theta {1\over{\bar{r}}} {{\cal Y}^{\bar{r}}}$ we find
an expression which is identical.
So finally we obtain
\be
{\cal J}= {{2^6}{e^{\gamma -2}}}\int {d^2}b\;d^2 z\;d^2 w\;
{{{{(1+|b|^2)^4}}{\bar{\vP}_0}(z,{\bar z}){\vP_0}(w,{\bar w})}\over
{{\epsilon^2}{(ln({\epsilon^2}))^3}{|c|^2}(z-b)({\bar w}-{\bar b})}}
\label{eq:express}
\ee
which is our proposed counterterm. It is divergent as
the zero instanton sector is approached from the one instanton sector as
${\epsilon^2}{(ln({\epsilon^2}))^3}\rightarrow 0$ as $\epsilon\rightarrow 0$.
However the $b$ integral is also infra-red divergent as a result of
our working on the plane so as to be able to use the results of
\cite{FFS}. This problem will be addressed in the next section.
\vskip0.5in
%%%%%%%%%%%%%%%%%%%%%%%%%%%%%%%%%%%%%%%%%%%%%%%%%%%%%%%%%%%%%%%%%%%%

\section{Conformal Invariance}

The infra-red divergence of the topological counterterm for large
values of the modulus $|b|$ can be approached by compactifying the
spacetime onto a spherical surface by performing a Weyl transformation
on the metric. This is
analogous to the method used in Yang-Mills Theory \cite{Paul} where a
similar problem arises. We will show that the result of doing this is
to render the integration over $b$ heavily damped for large $b$.

The metric on the plane $ds^2=d{\bar z}dz$ and on the sphere,
$ds^2=\Omega^2 d{\bar z}dz$ with $\Omega =1+{\bar z}z/h^2$, are related by
a Weyl transformation,
${g_{\mu\nu}}\rightarrow {e^{p(x)}}{g_{\mu\nu}}$
with $p=ln\, \Omega^2$. This can be built out of infinitessimal
transformations
${\del_p}{g_{\mu\nu}} = \delta p{g_{\mu\nu}}$. The field $w$ and quasi-ghost
$\xi$ are independent of the metric, and the classical action $S[w]$
is Weyl invariant. Thus the Green's Function moduli density
\be
{g(t)}=\int{\D W}\; {e^{-\m(S[w] + (\vs +
{\T^A}{\partial_A})({\xi^j}{F_j})\r)}}\;{\La(w)}
\ee
changes under the transformation ${\del_p}$ such that
\be
{\del_p}{g(t)}=\int{\D W}\;{e^{-S_{tot}}} \m[ \m(\int p{\cal M} - (\vs +
{\T^A}{\partial_A})({\xi^j}\;{\del_p}{F_j})\r)\;{\La(w)}
+{\del_p}{\La(w)}\r] \label{eq:wt}
\ee
where the first term is from the action of ${\del_p}$ on the volume element
and ${\cal M}$ is the Weyl anomaly density. The last term can be
removed provided ${\del_p}{\La(w)}=0$ and as $(\vs
+{\T^A}{\partial_A}){\La(w)}=0$ ($\vs$ doesn't act on $w$ and $w$ is
independent of the $t$) then
\be
{\del_p}{g(t)}=\int{\D W}\;{e^{-S_{tot}}}{\La(w)}\int p{\cal M} -
\int{\D
W}\;(\vs+{\T^A}{\partial_A}){e^{-S_{tot}}}({\xi^j}\;{\del_p}{F_j}){\La(w)}
\ee
However, the second term is linear in $\vP$ to first order in the
expansion of ${e^{-S_{tot}}}$, thus it does not contribute until next
to leading order. Also as  $\int \D
W\;{e^{-S_{tot}}}\;({\xi^j}{\del_p}{F_j})\;{\La(w)}$ is Grassmann odd,
\be
\int \D W\;\vs\;{e^{-S_{tot}}}\;({\xi^j}\del{F_j})\;{\La(w)} =0
\ee
So terms in (~\ref{eq:wt}) from the constraint piece in the action do
not contribute at the one loop level and we can write ${\del_p}{g(t)}$
as
\be
{\del_p}\int{\D W}\;{e^{-S[w]}}=\int{\D W}\;{e^{-S[w]}}\int \delta p{\cal M}
\label{eq:oneloop}
\ee
Thus the behaviour of the Green's Function on the sphere is governed, to
one loop, by the Weyl anomaly.

The calculation of the partition function involves the calculation of
$\det \DD$. However $\DD$ has dimensions of mass so it is convenient
to introduce an arbitrary parameter $\mu$ which also has dimensions of
mass.
To compensate for introducing $\mu$ the coupling constant becomes a
function of $\mu$, $k\rightarrow k(\mu)$.
This means that the coupling is no longer scale invariant as it is
classically.
Now an infinitessimal global scaleing of the metric
${g_{\mu\nu}}\rightarrow{g_{\mu\nu}} + \lambda{g_{\mu\nu}}$, can be
compensated by a shift in the mass-scale
${\del_\lambda}\mu=-{1\over 2}\lambda \mu$ and
${\del_\lambda}= {\del_\lambda}\mu\;{\partial\over{\partial \mu}}
=-{1\over 2}\lambda
\mu\;{\partial\over{\partial \mu}}$. Thus the action $S(w)$ is no
longer scale invariant either, since it contains the coupling. Using
$S(w)$ from (~\ref{eq:pact}) so
that now $S(w)={4\over{k(\mu)}} \int {d^2}x {\hat S}(w)$ then
\bq
{\del_\lambda} S(w)
&=&-2 \lambda\;\mu{\partial\over{\partial \mu}}
\m({1\over{k(\mu)}}\int {d^2}x {\hat S}(w) \r) \nonumber \\
&=&2 \lambda\;\mu\; {1\over{{k(\mu)}^2}}
{{\partial k(\mu)}\over{\partial \mu}} \int {d^2}x {\hat S}(w) \label{eq:hello}
\eq
The renormalisation group $\beta$-function for the
O(3) Sigma Model is given by $\beta= \mu{{\partial k}\over{\partial
\mu}}$, and is easily found to be $\beta={{k^2}\over{4\pi}}$.
Thus
\be
\int \lambda{\cal M}=-2 {\lambda\over{k^2}}\beta \int {d^2}x {\hat S}(w)
\ee
Hence
\be
{\cal M}=-2 {\beta\over{k^2}}{\hat S}(w)=-{1\over{2\pi}}{\hat S}(w)
\ee
up to total derivatives.
Now, knowing $\cal M$, the Green's Function can be evaluated on the
sphere using a position dependent scaleing $p$.

If ${g_h}(t)$ is the Green's Function moduli density
and ${\D_h} W$ the functional integral volume element when the sigma
model has as its space-time a sphere of radius $h$, then
\be
\int dt\;{g_h}(t)=\int dt\;{\D_h} W\;{e^{-S_{tot}}}{\La(w)}
\ee
if $\delta p=\del h \;{d\over{dh}} \ln {\Om^2}$ then to one loop
\bq
{\del_p}\int dt\;{g_h}(t)&=&\int dt\;{\D_h} W\;{e^{-S_{tot}}} \m(\int
\delta p{\cal M} \r){\La(w)}\nonumber \\
&=&\int dt\;{\D_h} W\;{e^{-S_{tot}}} \m(\int \del h \;\m({d\over{dh}}
\ln {\Om^2}\r){\cal M}\r){\La(w)}
\eq
so
\be
{d\over{dh}}\int dt\;{g_h}(t)=\int dt\;{\D_h} W\;{e^{-S_{tot}}}\m(\int
\m({d\over{dh}}\ln {\Om^2}\r){\cal M}\r){\La(w)}
\ee
Integrating with respect to $h$ from $h$ to infinity we obtain
\bq
\int dt\;{g_h}(t)
%&=&\int dt {\int^{\infty}_{h}} dh \;{\D_h} W\;{d\over{dh}}
%\m[{\exp\m({-S_{tot}+\int {d^2}x \ln \m({{\Om^2}\over 2}\r) {\cal
%M}}\r)}\r]
%{\La(w)}
%\nonumber \\
%&=&
=\int dt\;{\D_\infty} W\;\m[{\exp\m({-S_{tot}+\int {d^2}x \ln
\m({{\Om^2}\over 2}\r) {\cal M}}\r)}\r]{\La(w)}
\eq
The extra term involving the Weyl anomaly
suppresses the divergence due to the integral over $b$. Evaluating $\cal
M$ at the classical solution gives ${\hat S}(w)
=8{q^\prime}$ (see (~\ref{eq:sa}))
where ${q^\prime}$ is the topological charge density.
For small $a-b$ the charge becomes concentrated at $z=b$ so that this
density is approximately a delta-function
${q^\prime}\propto \del(z-b)$.
Thus the additional contribution to the action due to the Weyl anomaly is
\bq
\int {d^2} x\; \ln \m(\Om^2\r) {q^\prime} &=& \int {d^2} x\; \ln \m(
{2\over{1+{{z^2}\over{h^2}}}} \r)^2 \del(z-b) \nonumber \\
&=& \int {d^2} x\; \ln \m({2\over{1+{{b^2}\over{h^2}}}} \r)^2
\eq
which, for large $b$, will supply a strong damping factor for the
$b$-integration of Green's functions.
\vskip0.5in
%%%%%%%%%%%%%%%%%%%%%%%%%%%%%%%%%%%%%%%%%%%%%%%%%%%%%%%%%%%%%%%%%%%%%%%%
\section{Conclusions}

The moduli space integral of the $O(3)$ sigma model is divergent. To
control the divergence we introduce a cut-off in moduli-space. The
Green's functions then aquire a dependence on how the integration over
field configurations is split into a quantum and background piece.  It
is essential that physical quantities should not possess such a dependence.
To study this we split the field in the one instanton sector by
imposing a constraint on the quantum fluctuation and then varied
Green's functions with respect to this constraint. The resulting
`anomaly' is an integral over the boundary of moduli-space resulting from
the cut-off. It is in this region that one-instanton
configurations degenerate to configurations in the zero-instanton
sector. We found that the variation of the Green's function could be
expressed as the variation of a zero-instanton sector expression. This
provided us with a `counter-term' with which to cancel the
one-instanton moduli-space `anomaly'.

This `topological renormalisation' of the $O(3)$ sigma model is
very similar to that of bosonic string theory \cite{Paul}.
There is certainly scope for further work on this phenomenon for other
theories with pseudoparticle solutions.

\vskip0.5in

%%%%%%%%%%%%%%%%%%%%%%%%%%%%%%%%%%%%%%%%%%%%%%%%%%%%%%%%%%%%%%%%%%%%%%%%

\section*{Appendix A: Fluctuation operator and Greens Function}

When we introduce the quantum fluctuation $\vP(z,{\bar z})$
to the sigma model action of equation (~\ref{eq:sa}) we find that the
second term can be expressed as
\be
{8\over k}\int {d^2}\!x {{\partial_{\bar z}w}
{{\partial_z}{\bar w}}\over{(1 + |w|^2)^2}}
=(\vP,\DD\vP)
\ee
where
\be
\DD=-{{\rho^2}\over{\sqrt g}}{\partial_z}{\rho^{-2}}{\partial_{\bar
z}}
\ee
where ${\partial_z}={{\partial}\over{\partial{\bar z}}}$.
This is different to the $\DD$ found
in [2]. The difference is because the fluctuation itself is defined
differently in each case.
It will be useful to put $\DD$ in the form $\DD=T^{\dag}T$,
where $T^{\dag}$ is the adjoint of $T$ with respect to our inner
product (~\ref{eq:innprod}) so
$T={1\over{g^{1\over4}}}{\partial_{\bar z}}$ and
$T^{\dag}=-{{\rho^2}\over{\sqrt g}}{\partial_z}{g^{1\over4}}{\rho^{-2}}$.
The Green's Function of $\DD$ is the two-point Green's Function for
$\vP$, and satisfies
\be
\DD\I(x,y)={{\rho^2}\over{\sqrt g}}\;{\del^2}(x-y) - P(x,y) \label{eq:dgf}
\ee
where $P(x,y)$ is a projection operator
To find the form of $P(x,y)$ we look at the properties of $\I$. If
${\cal Z}(x)$ is a zero mode of $T$
(and again we use a dot notation to indicate an
inner product), then ${\bar{\cal Z}}(x).\DD\I(x,y) =
\m(T{\cal Z},T\I(x,y)\r)=0$
so ${\bar{\cal Z}}(x).P(x,y) = {\bar{\cal Z}}(y)$.
Also, $\vP$ is constrained by our choice of $F$ to be orthogonal to
$\psi$. So the two point function of $\vP$ and $\vP.\psi$ must vanish,
hence $\I(x,y).\psi(y)=0$ which
leads us to $P(x,y).\psi(y) = \psi(x)$, thus we can deduce
that
\be
P(x,y)={\psi_a}(x){{\bar m}^{-1}_{{\bar b}a}}{{\bar{\cal Z}}_{\bar
b}}(y)
\ee
We shall show that the one instanton sector
Green Function is
\be
\I(x,y)=-{1\over{\pi^2}}\int d^2 z \;
(1-P^{\dag}.){1\over{x-z}}{\rho^2}(z)
{1\over{{\bar z}-{\bar y}}}(1-.P)
\ee
with $P(x,y)$ as above, and $P^{\dag}(x,y)={{\cal Z}_b}(x){m^{-1}_{b{\bar a}}}
{{\bar \psi}_{\bar a}}(y)$. The dot notation, as above indicates an
inner product. So writing this out explicitly gives
\bq
\I(x,y)&=&-{1\over{\pi^2}}\int d^2 z \; \m[
{1\over{x-z}}{\rho^2}(z){1\over{{\bar z}-{\bar y}}} \r. \nonumber \\
& &\mbox{}-\int d^2 {x^{\prime}}\;P^{\dag}(x,{x^{\prime}})
{\sqrt g}\;{\rho^{-2}}({x^{\prime}}){1\over{{x^{\prime}}-z}}{\rho^2}(z)
{1\over{{\bar z}-{\bar y}}} \nonumber \\
& &\mbox{}-\int d^2 {y^{\prime}}\;{1\over{x-z}}{\rho^2}(z)
{1\over{{\bar z}-{\bar{y^{\prime}} }}}{\sqrt g}\;{\rho^{-2}}({y^{\prime}})
P({y^{\prime}},y) \nonumber \\
& &\mbox{}\m.-\int d^2{x^{\prime}}\;d^2 {y^{\prime}}\;P^{\dag}(x,{x^{\prime}})
{\sqrt g}\;{\rho^{-2}}({x^{\prime}}){1\over{{x^{\prime}}-z}}{\rho^2}(z)
{1\over{{\bar z}-{\bar{y^{\prime}} }}}{\sqrt g}\;{\rho^{-2}}({y^{\prime}})
P({y^{\prime}},y)\r]\nonumber
\eq
The deriavatives in $\DD$ act only on the $x$ variable, and as
${{\partial}\over{\partial{\bar x}}}{1\over{x-z}}={\pi}\;{\del^2}(x-z)$ then
\bq
{{\partial}\over{\partial{\bar x}}}\I(x,y)&=&-{1\over{\pi}}\int d^2 z \; \m[
{\del^2}(x-z){\rho^2}(z){1\over{{\bar z}-{\bar y}}}\r.\nonumber \\
& &\mbox{}\m.-\int d^2 {y^{\prime}}\;{\del^2}(x-z){\rho^2}(z)
{1\over{{\bar z}-{\bar{y^{\prime}}}}}{\sqrt g}\;{\rho^{-2}}({y^{\prime}})
P({y^{\prime}},y)\r]\nonumber \\
&=&-{1\over{\pi}}\int d^2 x {{{\rho^2}(x)}\over{{\bar x}-{\bar y}}}
+{1\over{\pi}}\int d^2 x\; d^2 {y^{\prime}}
{{{\rho^2}(x)}\over{{\rho^2}({y^{\prime}})}}
{{\sqrt g}\over{{\bar x}-{\bar{y^{\prime}}}}} P({y^{\prime}},y)
\eq
and
\bq
{{\partial}\over{\partial{x}}}\;{\rho^2}(x)\;{{\partial}\over{\partial{\bar
x}}}\;
\I(x,y)&=&-\int d^2 x \;{\del^2}({\bar x}-{\bar y}) \nonumber \\
& &\mbox{} +\int d^2 x\; d^2 {y^{\prime}} \;
{\del^2}({\bar x}-{\bar{y^{\prime}}})
{{\sqrt g}\over{{\rho^2}({y^{\prime}})}} P({y^{\prime}},y) \nonumber \\
&=&-1 + \int d^2 x\;{{\sqrt g}\over{{\rho^2}(x)}} P(x,y)
\eq
which gives us (~\ref{eq:dgf}).

In the zero instanton sector the Green Function becomes
\be
{\I_0}(x,y)=-{1\over{\pi^2}}\int d^2 z \;
(1-{\Pi^{\dag}}.){1\over{x-z}}{\rho^{2}_{0}}(z)
{1\over{{\bar z}-{\bar y}}}(1-.\Pi)
\ee
where $\Pi$ is the zero mode projector in the zero instanton sector
\be
\Pi f= {{\int\; {\sqrt g}\;f}\over{\int\; {\sqrt g}}} \;\;,\;\;
{\rho_{0}}=1+|c|^2
\ee
$\Pi$ and $P$ both project constant functions onto themselves since
these are zero-modes in both sectors. Hence $\Pi.P=\Pi$.
We need to find the
relationship between ${\I_0}(x,y)$ and ${\I}(x,y)$. First notice that
\be
(1-P^{\dag}.)\;{\I_0}(x,y)\;(1-.P)=-{1\over{\pi^2}}\int d^2 z \;
(1-P^{\dag}.){1\over{x-z}}{\rho^{2}_{0}}(z)
{1\over{{\bar z}-{\bar y}}}(1-.P)
\ee
Then for \e\/ small
\be
\rho={\rho_0} + O(\e)
\ee
so
\be
\I(x,y)=(1-P^{\dag}.)\;{\I_0}(x,y)\;(1-.P) \times \m(1+O(\e)\r)
\ee

\vskip0.5in
%%%%%%%%%%%%%%%%%%%%%%%%%%%%%%%%%%%%%%%%%%%%%%%%%%%%%%%%%%%%%%%%%%%%%%

\section*{Appendix B: K\"{a}hler Metric}

Here we give a brief review of the calculation to find the one
instanton K\"{a}hler Metric used above.
The metric tensor on the space of instanton parameters, found in
\cite{FFS}, is
\be
m_{AB} = \int {d^2}\!x \sqrt g
 \m({{\partial {\bar v}}\over{\partial t^A}}
{{\partial v}\over{\partial t^B}}\r){1\over{{(1+|v|^2)}^2}}
\ee
In a particular topological sector this can be written in terms of a
K\"{a}hler potential $\cal K$ since
\bq
{{\partial }\over{\partial t^{\al}}}
{{\partial}\over{\partial t^{\bar{\ba}}}}
\int {d^2}\!x\; \sqrt g\; \ln{(1+|v|^2)}
&=&{{\partial}\over{\partial t^{\bar{\ba}}}} \int {d^2}\!x\; \sqrt g\;
\m({1\over{{(1+|v|^2)}}}{\bar v}{{\partial v}\over{\partial
t^{\al}}}\r) \nonumber \\
&=&\int {d^2}\!x \sqrt g
\m({{\partial {\bar v}}\over{\partial t^{\bar{\ba}}}}
{{\partial v}\over{\partial t^{\al}}}\r){1\over{{(1+|v|^2)}^2}}
\eq
which has the same form as the metric tensor given above. The metric
$g_{\mu\nu}$ for the sphere $S^{2}_{phys}$ is
\be
g_{\mu\nu}=\del_{\mu\nu}\m(1+|z|^2\r)^{-2}
\ee
so
\be
\sqrt{g}=\m(1+|z|^2\r)^{-2}
\ee
thus
\be
m_{{\al}{\bar{\ba}}}={{\partial }\over{\partial t^{\al}}}
{{\partial}\over{\partial t^{\bar{\ba}}}}\; {\cal K} \nonumber
\ee
where
\be
{\cal K}=\int {d^2}\!x\;\m(1+|z|^2\r)^{-2}\;\ln{(1+|v|^2)}
\ee
which can be calculated. As $z=x+iy$ and the one instanton solution is
\be
v=c \m({{z-a}\over{z-b}}\r)
\ee
as before, then
\be
{\cal K}=\int  d x\; d y \; {(1+{|z|^2})}^{-2}\;
{\m[ln({|z-b|}^2 + {|c|}^2{|z-a|}^2)-ln({|z-b|}^2)\r]}
\ee
However, the second logarithm introduces a divergence in the limit
$z\rightarrow b$. This can be seen by looking at the double
differentiation of the second logarithm with respect to $b$ and $\bar
b$.
\be
{{\partial }\over{\partial b}}{{\partial }\over{\partial {\bar b}}}
ln({|z-b|}^2)={1\over\pi}{\del^2}(z-b)
\ee
This problem can be solved by noticing that as
\be
{1\over{{(1+|v|^2)}^2}}{{\partial v}\over{\partial b}}
{{\partial{\bar v}}\over{\partial{\bar b}}}=
{{{|c|}^2{|z-a|}^2}\over{{({|z-b|}^2 + {|c|}^2{|z-a|}^2)^2}}}=
{\partial\over{\partial b}}{\partial\over{\partial{\bar b}}}
\m({ln({|z-b|}^2+{|c|}^2{|z-a|}^2)}\r)
\ee
then we can ignore the second logarithm. We are left with
\be
{\cal K}=\int  d x\; d y \; {(1+{|z|^2})}^{-2}\;
{\m[ln({|z-b|}^2 + {|c|}^2{|z-a|}^2)\r]}
\ee
This integral can be solved by using two tricks to put the integrand
into an exponential form
\be
\int_0^{\infty} d{\al} \; \al \;
\exp \m(-\al(1+{|z|^2})\r)={(1+{|z|^2})}^{-2}
\ee
and
\be
ln(m)-ln(n)=\int_0^{\infty}{{d t}\over t}\; \m({e^{-mt}} -
{e^{-nt}}\r)
\ee
giving
\bq
{\cal K}&=&\int d \al\; d t\;d x\;d y {\alpha\over t}
\m[\exp\m(-({x^2}+{y^2})(\al+At)+2xtB+2ytC-Dt-\al \r)\r. \nonumber \\
& &\mbox{}{\hskip2.8in} -\m.\exp \m(-\al({x^2}+{y^2})-\al-t\r)\r]
\eq
where we have put
\be
A=(1+|c|^2),\;B=b_1 +{a_1}|c|^2,\;C=b_2 + {a_2}|c|^2,\; D=|b|^2 +
|a|^2|c|^2
\ee
so $a$ and $b$ have been split into their real and imaginary parts
($a=a_1 + ia_2$ etc.). The integrals over $x$ and $y$ can be done as
simple Gaussians, leaving
\be
{\cal K}=\int d \al\; d t\; {\alpha\over t}
\m[{\pi\over{\al+At}}\exp\m({{({B^2}+{C^2}){t^2}}\over{\al+At}}-(Dt+\al)\r)
-{\pi\over {\al}} \exp(-\al-t)\r]
\ee
To progress we change variables to $\lambda$ and $t$ where
$\al=\lambda t$,
$d \al=t\;d \lambda$ and
\bq
{\cal K}&=&\int d \lambda \; d t\; \pi
\m[{\lambda\over{\lambda+A}}\exp\m(t\m[{{({B^2}+{C^2})}\over{\lambda+A}}
-(D+\lambda)\r]\r)
-\exp\m(-t\m(\lambda+1\r)\r)\r] \nonumber \\
&=&\int d \lambda \;\pi \m[
{\lambda\over{{\lambda^2}+(A+D)\lambda+AD-({B^2}+{C^2})}}
-{1\over{\lambda+1}} \r]
\eq
The first term is just a standard integral of the form
\bq
\int_0^{\infty} {\T\over{\z +\eta\T +\h{\T^2}}}d \T &=&
{1\over{2\h}}ln(\z +\eta\T +\h{\T^2}) \nonumber \\
& &\mbox{} -{\eta\over{{2\h{{({\eta^2}-4\z\h)}^{1\over 2}}}}}
{ln{\m({{\eta+2\h\T-{({\eta^2}-4\z\h)}^{1\over 2}}
\over{\eta+2\h\T+{({\eta^2}-4\z\h)}^{1\over 2}}}\r)}}
\eq
The solution evaluated in
the limit $\lambda \rightarrow \infty$ is zero. So with
\be
G={{4(AD-{B^2}-{C^2})}\over{(A+D)^2}}
\ee
\be
{\cal K}={\pi\over 2} \m[
{1\over{{(1-G)^{1\over 2}}}}\;ln{\m({{1-{(1-G)^{1\over 2}}}
\over{1+{(1-G)^{1\over 2}}}}\r)}-ln(AD-{B^2}-{C^2}) \r]
\ee
To simplify this we can make the change of variables $a=b+r$, where
we are interested in the limit $r\rightarrow0$. In this case
$AD-{B^2}-{C^2}={|c|^2}{|a-b|^2}={|c|^2}{|r|^2}$, so $g$ is small and
{\cal K] can be expanded in powers of ${|r|^2}$. Terms of order $
{|\e|^4}$ can be dropped and
\bq
{\cal K}&=&\pi \m[ -ln\m(A+D\r) + {G\over 4}ln G - \m(1+{G\over 2}\r)ln
2 +{G\over 8}\r] \nonumber \\
& & \nonumber \\
&=& \pi \m[-lnS-{|c|^2 \over{S^2}}{(b{\bar{r}}+{\bar b}r)}
{{{|c|^2}{|r|^2}}\over{S^2}}
\m({1\over2}(1-2S)+2ln\m({{|c|^2}\over S}\r)+ln{|r|^2}\r)\r]
\eq
where $S=(1+{|b|^2})(1+{|c|^2})$.
This can now be differentiated to find $m_{{\al}{\bar{\ba}}}$.
Remember that we are now working with coordinates $\{t^{\al}\}=\{r,b,c\}$
and $\{t^{\bar{\ba}}\}=\{{\bar{r}},{\bar b}, {\bar c}\}$.
\be
m_{{\al},{\bar{\ba}}}=\pi\left(\matrix{ {\cal A}&{\cal B}&{\cal C}\cr
{{\cal B}^\dagger}&{\cal D}&{\cal E}\cr
{{\cal C}^\dagger}&{{\cal E}^\dagger}&{\cal F}\cr}\right)
\ee
where
\bq
{\cal A}&=&{|c|^2 \over{S^2}}\m[ {5\over 2}-S+{\cal V}\r] \nonumber \\
& & \nonumber \\
{\cal B}&=&-{\cal H}\m[S+{\bar{r}}b\m(5-S+2{\cal V}\r)\r] \nonumber
\\
& & \nonumber \\
{\cal C}&=&-{{c{\bar b}(1+{|b|^2})}\over{S^2}}
+{{c{\bar{r}}(1+{|b|^2})}\over{S^3}}\m[{7\over 2} - {{3{|c|^2}}\over
2} -S +(1-{|c|^2}){\cal V}\r] \nonumber \\
{\cal D}&=&-{1\over{(1+{|b|^2})^2}}+2{{(b{\bar{r}}+{\bar b}r)}{\cal H}}
\nonumber \\
{\cal E}&=&{c\over{S^2}}({{\bar b}^2}r-{\bar{r}}) \nonumber \\
{\cal F}&=&-{1\over{(1+{|c|^2})^2}}
\m[1+{{(b{\bar{r}}+{\bar b}r)}\over{S}}(1-{|c|^2})\r] \nonumber \\
{\cal H}&=&{{{|c|^2}{(1+{|c|^2})^2}}\over{S^3}}\;\;,\;\;
{\cal V}=ln{\m({{{|c|^4}|\e|^2}\over {S^2}}\r)} \nonumber \\
\eq
The inverse matrix given in (~\ref{eq:invm}) is
\be
{m^{-1}_{{\bar{\al}}\ba}}={1\over{m}} \left(\matrix{A&B&C\cr
                                              {B^\dagger}&D&E\cr
                                          {C^\dagger}&{E^\dagger}&F\cr}\right)
\ee
In the
limit $r \rightarrow 0$ the determinant is
\be
m={\rm det}\;m = {{|c|^2}\over{S^4}}\;ln{|r|^2}
\ee
and the \e\/ dependence of the components is
\bq
A&=&{A_1}+r{A_2}+{\bar{r}}{A_3}+{|r|^2}{A_4}+{r^2}{A_5}+
{{\bar{r}}^2}{A_6} \nonumber \\
B&=&{B_1}+r{B_2}+{\bar{r}}{B_3}+{|r|^2}{B_4}+{r^2}{B_5}+
{r \ln|r|^2}{B_6}+{|r|^2 \ln|r|^2}{B_7}+{r^2 \ln|r|^2}{B_8}
\nonumber \\
C&=&{C_1}+r{C_2}+{\bar{r}}{C_3}+{|r|^2}{C_4}+{r^2}{C_5}+
{r \ln|r|^2}{C_6}+{|r|^2 \ln|r|^2}{C_7}+{r^2 \ln|r|^2}{C_8}
\nonumber \\
D&=&{\ln|r|^2}{D_0}+{D_1}+r{D_2}+{\bar{r}}{D_3}+{|r|^2}{D_4}+
{{\bar{r}} \ln|r|^2}{D_5} \nonumber \\
& &\mbox{}{\hskip2in}
+{r \ln|r|^2}{D_6}+{|r|^2 \ln|r|^2}{D_7}+{|r|^2 (\ln|r|^2)^2}{D_8}
\nonumber \\
E&=&{E_1}+r{E_2}+{\bar{r}}{E_3}+{|r|^2}{E_4}+{{\bar{r}} \ln|r|^2}{E_5}+
{r \ln|r|^2}{E_6}+{|r|^2 \ln|r|^2}{E_7}+{|r|^2 (\ln|r|^2)^2}{E_8}
\nonumber \\
F&=&{\ln|r|^2}{F_0}+{F_1}+r{F_2}+{\bar{r}}{F_3}+{|r|^2}{F_4}+
{{\bar{r}} \ln|r|^2}{F_5} \nonumber \\
& &\mbox{}{\hskip2in}
{r \ln|r|^2}{F_6}+{|r|^2 \ln|r|^2}{F_7}+{|r|^2 (\ln|r|^2)^2}{F_8}
\eq
where the $A_i$'s, $B_i$'s, $C_i$'s, $D_i$'s, $E_i$'s and $F_i$'s are
all just functions of $b$ and $c$. The term used in (~\ref{eq:divterm1})
is
\be
A={1\over{{\cal S}^2}}
\ee

%%%%%%%%%%%%%%%%%%%%%%%%%%%%%%%%%%%%%%%%%%%%%%%%%%%%%%%%%%%%%%%


\begin{thebibliography}{88}

\bibitem{Raj} R. Rajaraman, {\em Solitons and Instantons},
(North-Holland, Amsterdam, 1989)

\bibitem{God+Man} P. Goddard and P. Mansfield, Rep. Prog. Phys. 49
(1986) 725

\bibitem{AMP} A. M. Perelomov,  Physica 4D (1981) 1 and  Phys. Rep. 146
No.3 (1987) 135

\bibitem{Bel+Poly} A. A. Belavin and A. M. Polyakov,  JETP Lett Vol.22
No.10 (1975) 245

\bibitem{FFS} V.A. Fateev, I.V. Frolov and A.S. Schwarz,  Nucl. Phys. B154
(1979) 1

\bibitem{BRS} C. Becchi, A. Rouet and R. Stora,  Phys. Lett. B52 (1974)
344

\bibitem{Paul} P. Mansfield,  Nucl. Phys. B416 (1994) 205

\bibitem{D+P} E. D'Hoker and D. H. Phong,  Rev. Mod. Phys. 60 (1988) 917

\bibitem{C+G} J. Cervero and C. Gomez,  Phys. Lett. B104 (1981) 467

\bibitem{Adda} A. D'Adda, M. Luscher and P. DiVecchia,  Nuc. Phys. B146
(1978) 63

\bibitem{BPST} A. A. Belavin, A. M. Polyakov, A.S. Schwarz and Yu. S.
Tyupkin,  Phys. Lett. 59B (1975) 85

\bibitem{Jack} R. Jackiw,  Rev. Mod. Phys. 52 (1980) 4

\bibitem{FP} L. D. Faddeev and V. N. Popov,  Phys. Lett. B25 (1967) 29

\bibitem{Poly} A. M. Polyakov,  Phys. Lett. B103 (1981) 207

\bibitem{t'H} G. t'Hooft,  Phys. Rev. D14 (1976)3432

\bibitem{Osb} H. Osborn,  Ann. Phys. 135 (1981) 373

\bibitem{Nak} M. Nakahara, {\em Geometry, Topology and Physics}, (Adam
Hilger, Bristol, 1990)

\bibitem{Woj} W. J. Zakrzewski, {\em Low Dimensional Sigma Models}, (Adam
Hilger, Bristol, 1989)
\end{thebibliography}
\end{document}